\documentclass[runningheads]{llncs}
\usepackage[T1]{fontenc}
\usepackage{graphicx}
\usepackage{cite}
\usepackage{amsmath,amssymb,amsfonts}
\usepackage{graphicx}
\usepackage{textcomp}
\usepackage{xcolor}

\usepackage{subcaption}
\usepackage{algorithm}
\usepackage{algpseudocode}
\usepackage{algorithmicx}
\usepackage{array}
\usepackage{booktabs}
\usepackage{multirow} 
\usepackage{placeins}
\usepackage{wrapfig}
\newtheorem{assumption}{Assumption}
\usepackage{orcidlink}
\hypersetup{hidelinks}

\begin{document}

\title{SuperSFL: Resource-Heterogeneous Federated Split Learning with Weight-Sharing Supernet}
\titlerunning{SuperSFL}

\author{
Abdullah Al Asif\inst{1}\orcidlink{0000-0002-0389-931X} \and
Sixing Yu\inst{2}\orcidlink{0000-0002-2415-566X} \and
Juan Pablo Muñoz\inst{3}\orcidlink{0000-0002-5901-4023} \and
Arya Mazaheri\inst{4,5}\orcidlink{0000-0002-5671-4710} \and
Ali Jannesari\inst{1}\orcidlink{0000-0001-8672-5317}
}

\authorrunning{A. A. Asif et al.}

\institute{
Department of Computer Science, Iowa State University, Ames, IA, USA \\
\email{\{aaasif, jannesari\}@iastate.edu}
\and
Microsoft, USA \\
\email{sixingyu@microsoft.com}
\and
Maro Systems, USA \\
\email{pablo.munoz@maro-systems.com}
\and
Technical University of Darmstadt, Darmstadt, Germany \\
\and
PanocularAI \\
\email{arya@panocular.ai}
}




\maketitle              
\begin{abstract}
Real-world edge devices differ widely in memory and connectivity, so they cannot all split a shared model at the same layer. Allowing heterogeneous split depths, however, produces conflicting gradient signals that destabilize training.
SuperSFL addresses this challenge with a single weight-sharing supernetwork whose prefix each client executes up to a depth matching its resources. To stabilize optimization across these mismatched splits, we introduce Three-Phase Gradient Fusion (TPGF). TPGF merges client- and server-side gradients using loss- and depth-aware weights that follow from a bias–variance analysis of the two supervision sources. A lightweight local classifier further enables clients to continue training locally during server outages and seamlessly reintegrate their updates when connectivity returns.
Experiments on CIFAR-10 and CIFAR-100 show that SuperSFL converges up to 4.4× faster while reducing communication and energy consumption by up to 4× and 4.6×, respectively, compared with prior split and heterogeneous FL methods.
\end{abstract}


\section{Introduction}
\label{sec:intro}

Federated Learning (FL) enables collaborative model training across
multiple devices without transferring raw data to a central server
\cite{mcmahan2017communication}. This property makes FL attractive for
privacy-sensitive applications such as mobile sensing, healthcare, and
intelligent IoT systems \cite{kairouz2021advancesopenproblemsfederated,fedprox}.
However, real-world edge environments are highly heterogeneous.
Devices differ in memory capacity, computational capability, network
bandwidth, and connectivity reliability. These differences often
violate the assumptions of standard FL pipelines and can slow down
training or prevent weaker devices from participating effectively
\cite{fedprox,kairouz2021advancesopenproblemsfederated}.

\begin{figure}[t]
\centering
\includegraphics[width=\textwidth]{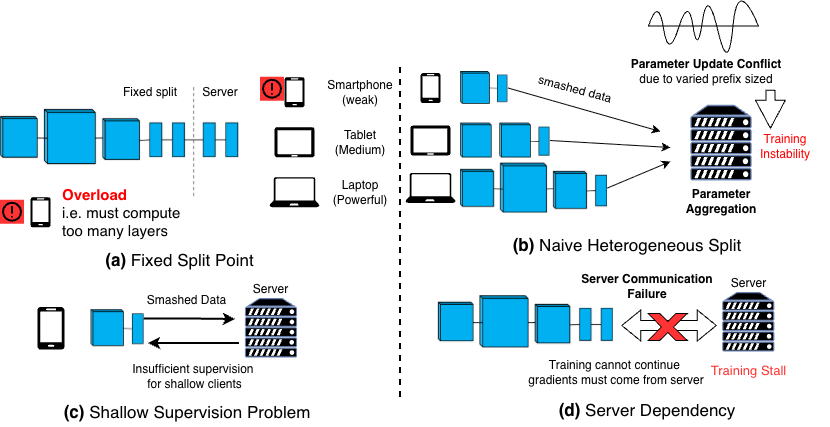}
\caption{Challenges of split learning under heterogeneous client resources}
\label{fig:problem}
\end{figure}

Split Learning (SL) addresses this limitation by dividing the model
between the client and the server \cite{vepakomma2018split}. Only a
prefix of the network is executed on the client, while the remaining
layers are computed on the server. This design reduces memory and
compute requirements on edge devices and allows resource-constrained
clients to participate in collaborative training. Split Federated
Learning (SFL) extends this idea by combining split learning with
federated aggregation so that multiple clients can train in parallel
while preserving data locality \cite{thapa2020splitfed}. Despite these advantages, heterogeneous environments introduce new
challenges for split learning. In practice, different devices often
require different split depths depending on their available resources. Allowing such heterogeneous
splits, however, creates inconsistent gradient signals across the
shared encoder. Clients interact with different portions of the model,
which can lead to unstable optimization and degraded convergence.
Existing SFL approaches typically assume fixed split boundaries or
rely on static partitioning strategies that do not address this issue
\cite{thapa2020splitfed}. Several heterogeneous FL methods attempt to address resource variability
by allowing clients to train models of different sizes
\cite{diao2021heterofl,caldas2018leaf}. These approaches focus on
parameter heterogeneity but operate within a single-endpoint training
topology. As a result, they do not resolve the cross-boundary
optimization challenges that arise when clients train different
segments of a shared representation pipeline in split learning.

Figure~\ref{fig:problem} illustrates the key challenges that emerge in
heterogeneous split learning environments. First, fixed split
configurations may overload weak devices while under-utilizing more
capable ones. Second, na\"ively allowing heterogeneous split depths can
produce conflicting gradient updates because clients train different
prefixes of the network. Third, shallow clients may receive insufficient
supervision when most of the learning signal originates from deeper
layers. Finally, many split learning pipelines depend heavily on the
server, which can stall training when connectivity becomes unreliable. To address these challenges, we propose \textbf{SuperSFL}, a resource-aware
split learning framework designed for heterogeneous edge systems.
SuperSFL trains a shared weight-sharing supernetwork while allowing
each client to use a prefix subnetwork that matches its computational
capacity. This design enables flexible split depths while maintaining
a unified parameter space for collaborative training. To stabilize optimization across heterogeneous split configurations,
we introduce Three-Phase Gradient Fusion (TPGF). TPGF combines
client-side and server-side gradients using loss-aware and
depth-aware weighting to align updates originating from different
split boundaries. In addition, SuperSFL includes a lightweight
client-side classifier that allows clients to continue learning
locally during temporary server outages and reintegrate updates once
connectivity is restored. This paper makes the following contributions:

\begin{itemize}
    \item We propose a weight-sharing supernetwork framework that
    enables clients to train prefix subnetworks with different split
    depths while maintaining a shared parameter space.

    \item We introduce Three-Phase Gradient Fusion (TPGF), a loss- and
    depth-aware optimization mechanism that stabilizes training across
    heterogeneous split configurations.

    \item We design a fault-tolerant split learning pipeline that
    reduces server dependency and allows training to continue during
    intermittent connectivity.
\end{itemize}

\section{Background and Related Work}
\label{sec:background}
Federated Averaging~\cite{mcmahan2017communication} established the blueprint for 
collaborative training without sharing raw data, but it assumes every client trains 
the same model on a reliable connection. Neither holds in practice. Split 
Learning~\cite{vepakomma2018split} and its federated extension 
SFL~\cite{thapa2020splitfed} help by offloading most of the network to the server, 
yet they inherit the same fixed-split assumption that breaks as soon as clients 
differ in memory or bandwidth~\cite{hukkeri2025split}. DFL~\cite{samikwa2024dfl} 
adapts split points through clustering but does not address what happens when 
different clients backpropagate through different portions of a shared encoder.

Model-heterogeneous FL takes a different angle: give each client a smaller version 
of the model~\cite{diao2021heterofl,ye2023heterogeneous}. This works for 
single-endpoint training but ignores the cross-boundary gradient conflicts that 
split architectures introduce. Once-for-All~\cite{cai2020once} and related federated 
supernet work~\cite{kim2022supernet} show that weight sharing across subnetworks is 
tractable, and personalized NAS methods like PerFedRLNAS~\cite{prefednas} push this 
further, but they train full models rather than split prefixes. On the fault 
tolerance side, asynchronous FL~\cite{morell2022dynamic} and 
FTFormer~\cite{zhang2025ftformer} handle dropout and layer offloading respectively, 
yet server dependency in split learning remains largely unsolved. SuperSFL sits at 
the intersection of all three problems.

\section{Methodology}

\subsection{The SuperSFL Framework}

\begin{figure}[tb]
\centering
\includegraphics[width=0.9\textwidth]{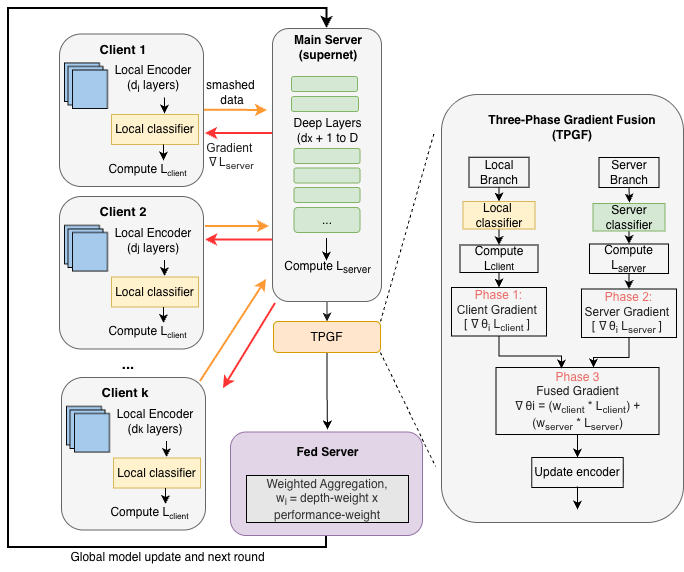}
\caption{SuperSFL architecture and end-to-end training workflow. 
(\textbf{Left}) Clients run prefix encoders of varying depths $d_i$ with lightweight auxiliary classifiers and forward smashed data to the Main Server (supernet layers). 
The TPGF module (orange) fuses client- and server-side gradients before the Fed Server (purple) performs depth- and performance-aware aggregation. 
(\textbf{Right}) Three-Phase Gradient Fusion in detail: Phase 1 computes the client-side gradient from the local classifier; Phase 2 recovers the server-side gradient via backpropagation through the client encoder; Phase 3 combines both with loss- and depth-aware weights.}
\label{fig:superSFL}
\end{figure}

SuperSFL organizes collaborative training around a single
weight-sharing supernetwork
$\theta = \{\theta_1, \ldots, \theta_L\}$ shared across all
participants.
Rather than forcing every client to split the model at the same
layer, each client executes a prefix subnetwork whose depth matches
its available resources.
Figure~\ref{fig:superSFL} shows the resulting workflow: clients send
smashed data to the Main Server, which completes the forward pass
and returns gradient signals; the TPGF module then merges local and
server-side gradients before the Fed Server aggregates updates into
the global supernetwork.
The four components below each address a distinct failure mode of
standard split learning under device heterogeneity.

\paragraph{Resource-Aware Subnetwork Allocation.}
Assigning split depths na\"{i}vely leads to two opposing problems:
resource-constrained devices overloaded with deep prefixes, and
high-latency devices incurring excessive communication costs.
To balance these, each client reports its available memory $m_i$
and round-trip latency $\mathrm{lat}_i$ once at initialization.
SuperSFL then assigns a prefix depth according to
\begin{equation}
d_i = \min\!\left(
  \left\lfloor
    \alpha\, m_i
    + \beta\,\frac{\mathrm{lat}_\mathrm{max} - \mathrm{lat}_i}
                  {\mathrm{lat}_\mathrm{max} - \mathrm{lat}_\mathrm{min}
                   + \varepsilon}
  \right\rfloor,\;
  L - 1
\right), \quad d_i \geq 1,
\label{eq:depth}
\end{equation}
where $\alpha = 0.5$ layers/GB scales with memory and $\beta = 4$
scales with relative latency rank.
The floor function ensures $d_i$ is a valid layer index.
The two terms trade off compute capacity against communication cost:
a client with ample memory but poor connectivity receives a shallower
prefix than one with the same memory but low latency.
Because each prefix is a contiguous slice of the shared supernetwork,
parameters at every depth are structurally identical across clients,
keeping aggregation well-defined regardless of how many layers each
client trains.

\paragraph{Three-Phase Gradient Fusion (TPGF).}
The core optimization challenge in heterogeneous split learning is
that clients with different split depths produce gradient updates
covering different portions of the shared encoder.
Relying only on server-side gradients breaks down under intermittent
connectivity, while relying only on the local classifier provides
insufficient supervision for shallow encoders.
TPGF addresses this by running two gradient branches in parallel and
combining them into a single update, as shown in the right panel of
Figure~\ref{fig:superSFL}.

\textit{Phase~1 --- Local supervision.}
Client $i$ maps its input $x_i$ through the encoder $f_{\theta_i}$
to produce smashed data $z_i^c = f_{\theta_i}(x_i)$.
A lightweight local classifier $h_{\varphi_i}$ then produces a
prediction $\hat{y}_i$, giving the local cross-entropy loss
$\mathcal{L}_\mathrm{client} = \mathrm{CE}(\hat{y}_i, y_i)$.
The classifier parameters $\varphi_i$ are updated by gradient
descent, and the encoder gradient $\nabla_{\theta_i}\mathcal{L}_\mathrm{client}$ 
is clipped to $\ell_2$-norm $\leq 0.5$~\cite{pascanu2013difficulty} 
to guard against instability on low-precision hardware.

\textit{Phase~2 --- Server supervision.}
The smashed data $z_i^c$ is transmitted to the Main Server along
with the ground-truth label $y_i$.
The server computes
$\mathcal{L}_\mathrm{server}
  = \mathrm{CE}(h_{\varphi_s}(f_{\theta_s}(z_i^c)),\, y_i)$
and updates its own parameters.
It then returns $\nabla_{z_i^c}\mathcal{L}_\mathrm{server}$, from
which the client recovers
$\nabla_{\theta_i}\mathcal{L}_\mathrm{server}$ by backpropagating
through its local encoder.
Transmitting $y_i$ introduces negligible overhead: a single integer
of $\lceil\log_2 C\rceil$ bits versus the full smashed-data tensor
already sent in standard split learning.
Regarding privacy, a class label conveys only coarse semantic
information and cannot reconstruct $x_i$, preserving split
learning's core guarantee; for stricter requirements, $y_i$ can be
protected via label differential privacy~\cite{ghazi2021deep}
without modifying TPGF.

\textit{Phase~3 --- Fusion and encoder update.}
The client now holds two gradient estimates for the shared encoder
parameters: one from shallow local supervision and one from deep
server supervision.
Let $d_s = L - d_i$ denote the number of layers processed on the
server.
Fusion proceeds in two stages.
First, a \emph{direction weight} $w_\mathrm{client}$ determines
which gradient signal to trust, based on current supervision
quality:
\begin{equation}
  w_\mathrm{client} =
    \frac{\bigl(\mathcal{L}_\mathrm{client}+\varepsilon\bigr)^{-1}}
         {\bigl(\mathcal{L}_\mathrm{client}+\varepsilon\bigr)^{-1}
          + \bigl(\mathcal{L}_\mathrm{server}+\varepsilon\bigr)^{-1}},
  \qquad
  w_\mathrm{server} = 1 - w_\mathrm{client},
  \label{eq:direction}
\end{equation}
where $\varepsilon = 10^{-8}$.
The inverse-loss weighting assigns higher trust to whichever branch
currently produces lower loss, and naturally increases
$w_\mathrm{client}$ for shallower clients whose larger $d_s$ tends
to degrade server supervision.
Second, a \emph{coverage scalar} $\rho_i$ normalizes the magnitude
of the fused update in proportion to the fraction of the
supernetwork the client actually trained:
\begin{equation}
  \rho_i = \frac{d_i}{d_i + d_s}.
  \label{eq:coverage}
\end{equation}
The encoder is then updated as
\begin{equation}
  \nabla\theta_i =
    \rho_i \Bigl(
      w_\mathrm{client}\,\nabla_{\theta_i}\mathcal{L}_\mathrm{client}
      + w_\mathrm{server}\,\nabla_{\theta_i}\mathcal{L}_\mathrm{server}
    \Bigr),
  \qquad
  \theta_i \;\leftarrow\; \theta_i - \eta\,\nabla\theta_i.
  \label{eq:fusion}
\end{equation}
Fusion is applied only to the shared encoder parameters; the local
classifier head remains decoupled from the global supernetwork.

\paragraph{Fault-Tolerant Client-Side Continuation.}
Standard SFL stalls completely when the server is unreachable,
because gradients can only flow from the server back to the client.
SuperSFL avoids this through the local classifier already present
for Phase~1: during a server outage, this classifier doubles as a
standalone training head.
If no server response arrives within a timeout threshold, the client
switches to fallback mode and continues updating $\theta_i$ and
$\varphi_i$ using only the Phase~1 gradient, so local training
never stops.
When connectivity is restored, the client pulls the latest global
parameters and returns to full TPGF-based training; the fallback
updates are incorporated into the next aggregation round through
the weighting scheme described below.
This design adds no extra parameters and no additional communication
overhead to the normal training path.

\paragraph{Aggregation Across Heterogeneous Subnetworks.}
Clients submit encoder updates of different depths and, after server
outages, updates trained under different supervision levels.
Standard FedAvg~\cite{mcmahan2017communication} handles neither case, because it
assumes uniform model size and consistent gradient quality.
Instead, the Fed Server assigns each client an aggregation score
\begin{equation}
\tilde{w}_i \;=\;
  \frac{d_i}{\sum_j d_j}
  \cdot
  \frac{\bigl(\mathcal{L}_i + \varepsilon\bigr)^{-1}}
       {\sum_j \bigl(\mathcal{L}_j + \varepsilon\bigr)^{-1}},
\label{eq:aggr-score}
\end{equation}
that jointly reflects encoder depth and loss quality.
Since $\tilde{w}_i$ is a product of two distributions it does not
sum to 1 in general; it is therefore normalized into proper
aggregation weights:
\begin{equation}
w_i \;=\; \frac{\tilde{w}_i}{{\displaystyle\sum_j \tilde{w}_j}}.
\label{eq:aggr-weight}
\end{equation}
Layer-wise averaging then proceeds for each shared encoder layer
$\ell$ over all clients whose prefix reaches that layer, with a
consistency regularizer anchoring the result to the server
parameters:
\begin{equation}
\bar{\theta}^{\ell} \;=\;
  \frac{{\displaystyle\sum_i w_i\,\theta_i^{\ell}}
        \;+\; \lambda\,\theta_s^{\ell}}
       {{\displaystyle\sum_i w_i} \;+\; \lambda},
\label{eq:aggr-layer}
\end{equation}
where $\lambda = 0.01$ throughout.
The regularizer $\lambda\theta_s^\ell$ acts as a Gaussian prior
centered at the server parameters, equivalent to FedProx-style
proximal regularization~\cite{fedprox} applied layer-wise, and
prevents the global model from drifting when a subset of clients
trained entirely without server gradients.

\subsection{Bias--Variance Optimal Fusion in TPGF}

The two branches in TPGF produce gradient estimates with different
bias and variance profiles.
Shallow local supervision is biased but low-variance; server
supervision is better aligned with the global objective but becomes
noisier as the missing depth $d_s = L - d_i$ increases.
We formalize this as a multi-source gradient estimation problem and
show that the inverse-loss factor in Eq.~\eqref{eq:direction}
approximates the bias--variance optimal fusion weight.
The coverage scalar $\rho_i$ addresses the orthogonal issue of
proportional network coverage and is revisited in
Remark~\ref{rem:tpgf}.

Let $\nabla F_i(\theta)$ denote the true client gradient for the
shared encoder, and let the global objective be
$F(\theta)=\sum_{i=1}^{K}p_i F_i(\theta)$ with
$p_i = d_i/\sum_j d_j$.
The local and server branches compute
\begin{equation*}
  g_i^c = \nabla F_i(\theta) + b_i^c + \varepsilon_i^c,
  \qquad
  g_i^s = \nabla F_i(\theta) + b_i^s + \varepsilon_i^s,
\end{equation*}
where $b_i^c$, $b_i^s$ are deterministic bias terms and
$\varepsilon_i^c$, $\varepsilon_i^s$ are zero-mean stochastic
components.
We work under the following assumptions.

\begin{assumption}[Bounded bias]
\label{ass:bias}
$\|b_i^c\| \leq \beta_i^c$ and $\|b_i^s\| \leq \beta_i^s$,
with larger bias under shallow supervision due to task mismatch
and depth-dependent model-path mismatch.
\end{assumption}

\begin{assumption}[Bounded stochastic variance]
\label{ass:var}
$\mathbb{E}\|\varepsilon_i^c\|^2 \leq \sigma_i^{c,2}$ and
$\mathbb{E}\|\varepsilon_i^s\|^2 \leq \sigma_i^{s,2}
  = \sigma_0^2 + \gamma\tfrac{d_s}{L}$
for $\gamma > 0$ ~\cite{pennington2018emergence}.
This follows from the chain rule: backpropagating through $d_s$
additional server-side layers multiplies the gradient signal by
$d_s$ Jacobians each with bounded spectral norm, causing variance
to accumulate linearly with $d_s/L$.
\end{assumption}

\begin{assumption}[Bounded bias correlation and uncorrelated noise]
\label{ass:corr}
$|\langle b_i^c,\, b_i^s \rangle| \leq \delta$ for some
$\delta \geq 0$, and
$\mathbb{E}\langle\varepsilon_i^c,\,\varepsilon_i^s\rangle = 0$.
The case $\delta = 0$ recovers strict orthogonality; $\delta$ is
expected to be small as branch supervision depths diverge.
\end{assumption}

\begin{proposition}[Bias--Variance Optimal Fusion]
\label{prop:tpgf}
Under Assumptions~\ref{ass:bias}--\ref{ass:corr},
the fused gradient
$\tilde{g}_i = w_c\,g_i^c + (1-w_c)\,g_i^s$
satisfies
\begin{equation}
  \mathbb{E}\|\tilde{g}_i - \nabla F_i(\theta)\|^2
  \;\leq\;
  w_c^2\!\left(\|b_i^c\|^2 + \sigma_i^{c,2}\right)
  +
  (1-w_c)^2\!\left(\|b_i^s\|^2 + \sigma_i^{s,2}\right)
  + \frac{\delta}{2}.
  \label{eq:bv}
\end{equation}
The dominant terms are minimized over $w_c \in [0,1]$ at
\begin{equation}
  w_c^{*}
  =
  \frac{\|b_i^s\|^2 + \sigma_i^{s,2}}
       {\|b_i^s\|^2 + \sigma_i^{s,2}
        + \|b_i^c\|^2 + \sigma_i^{c,2}},
  \label{eq:wopt}
\end{equation}
which increases monotonically with $d_s$.
Since $\delta/2$ is independent of $w_c$ it does not shift the
minimizer, and shallower clients have a higher optimal local weight.
\end{proposition}

\begin{proof}
Substitute the gradient decompositions and take expectation.
The cross-noise term vanishes by Assumption~\ref{ass:corr}.
For the bias cross-term, Assumption~\ref{ass:corr} gives
$2w_c(1-w_c)\langle b_i^c, b_i^s\rangle
  \leq 2w_c(1-w_c)\delta \leq \tfrac{\delta}{2}$,
since $w_c(1-w_c) \leq \tfrac{1}{4}$ for all $w_c \in [0,1]$,
yielding Eq.~\eqref{eq:bv}.
Setting $\mathrm{d}\phi/\mathrm{d}w_c = 0$ on the dominant terms
$\phi(w_c)
  = w_c^2(\|b_i^c\|^2+\sigma_i^{c,2})
  + (1-w_c)^2(\|b_i^s\|^2+\sigma_i^{s,2})$
yields Eq.~\eqref{eq:wopt}; the second derivative is strictly
positive, confirming a unique global minimum on $[0,1]$.
\end{proof}

\begin{remark}[Interpretation of TPGF weights]
\label{rem:tpgf}
Proposition~\ref{prop:tpgf} governs $w_\mathrm{client}$ in
Eq.~\eqref{eq:direction}: the optimal $w_c^*$ increases with $d_s$,
so shallower clients trust their local gradient more as server
supervision degrades.
The coverage scalar $\rho_i$ in Eq.~\eqref{eq:coverage} addresses
a separate concern---update magnitude---scaling the fused gradient
by the fraction of the supernetwork the client trained and
preventing shallow clients from dominating aggregation.
Since $\delta/2$ in Eq.~\eqref{eq:bv} is constant in $w_c$ it does
not shift the minimizer, and the approximation in
Eq.~\eqref{eq:direction} remains valid.
The two factors are therefore complementary: $w_\mathrm{client}$
selects gradient direction; $\rho_i$ scales the step.
\end{remark}
\section{Results}
\label{sec:results}

We evaluate SuperSFL on CIFAR-10 and CIFAR-100 using two backbones: ViT-B/16~\cite{dosovitskiy2021an} 
and ResNet-18~\cite{he2016deep}. We simulate device heterogeneity in a controlled 
setting to isolate algorithmic effects and ensure fair comparison. Following standard 
practice in federated and split learning studies~\cite{mcmahan2017communication,
thapa2020splitfed,diao2021heterofl}, each client is assigned synthetic memory and 
latency values according to Eq.~\eqref{eq:depth}. All methods are trained under the 
same heterogeneous profile, so differences in performance reflect optimization 
strategy rather than environmental bias.

Client models are prefix subnetworks of a shared supernetwork following prior 
split-learning designs~\cite{cai2020once,vepakomma2018split}. Experiments involve 
50 and 100 clients with non-i.i.d.\ data generated by a Dirichlet distribution 
($\alpha = 0.5$). We compare SuperSFL against SplitFed~\cite{thapa2020splitfed}, 
DFL~\cite{samikwa2024dfl}, and HeteroFL~\cite{diao2021heterofl}. All methods use 
Adam ($\eta{=}10^{-3}$, weight decay $10^{-4}$, batch size 64) for 100 
communication rounds. Baseline hyperparameters were tuned independently by grid 
search over learning rate $\in \{5{\times}10^{-4},\,10^{-3},\,2{\times}10^{-3}\}$ 
and local epochs $\in \{1, 2, 3\}$ using the same validation split as SuperSFL.

\vspace{-1em}
\subsection{Overall Convergence Performance}

\begin{figure}[t]
\centering
\includegraphics[width=\textwidth]{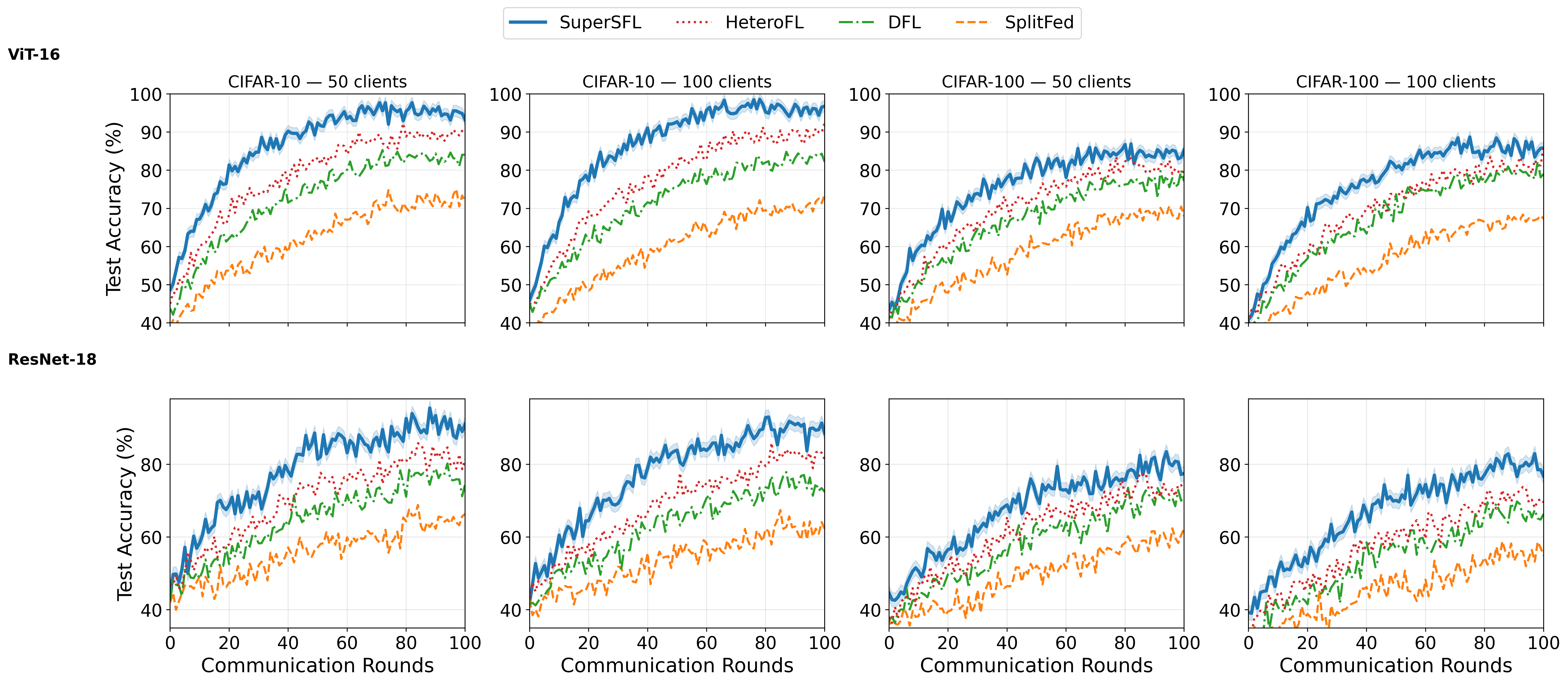}
\caption{Convergence curves (test accuracy vs. communication rounds) on CIFAR-10 and CIFAR-100 with 50 and 100 clients (Dirichlet $  \alpha=0.5  $) for SuperSFL, SplitFed, DFL, and HeteroFL. Each curve reports mean $  \pm  $ standard deviation over three independent runs.}
\label{fig:model_comparison}
\end{figure}

\begin{figure}[t]
\centering
\includegraphics[width=\columnwidth]{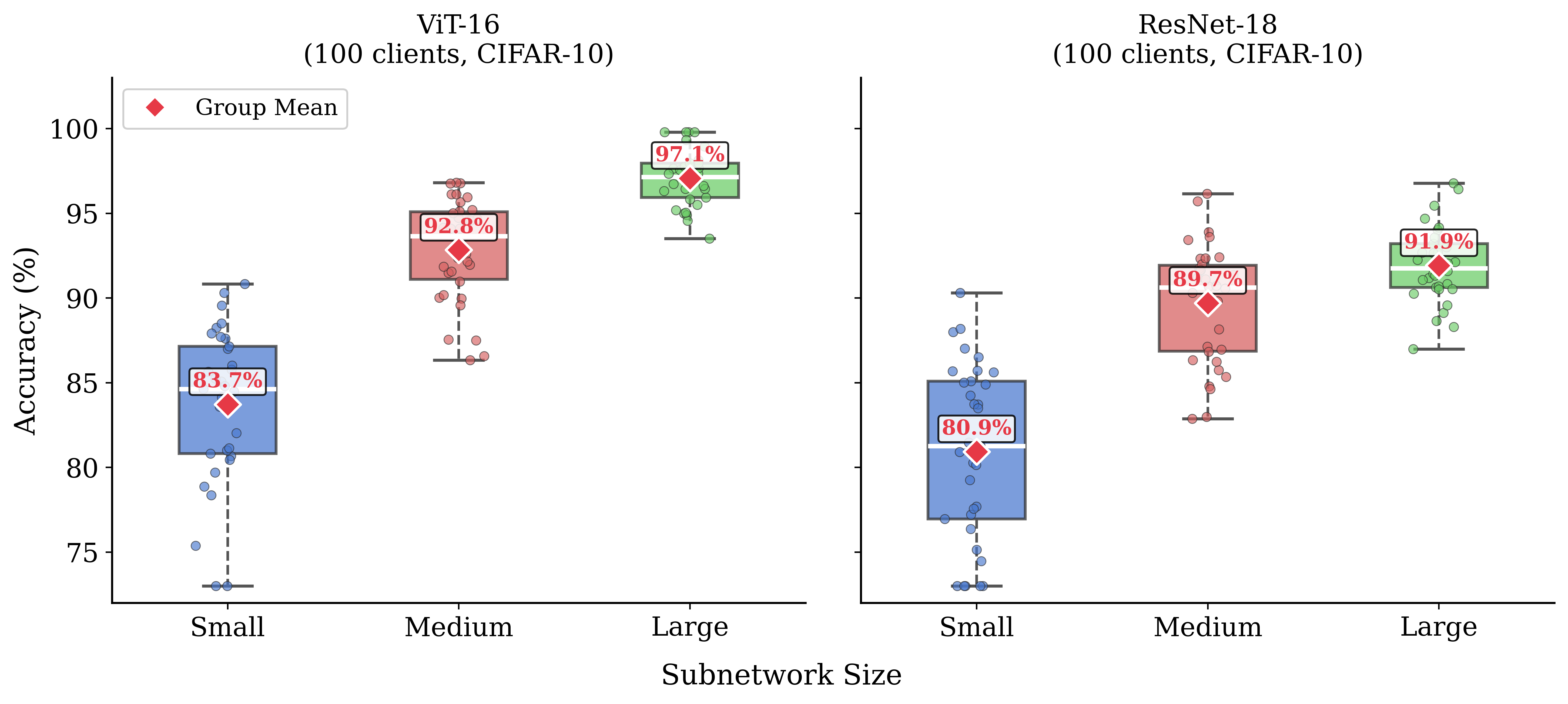}
\caption{Per-client final accuracy grouped by assigned subnetwork size. Red diamonds mark group means.}
\label{fig:client-wise-acc}
\end{figure}

\vspace{-0.5em}

Figure~\ref{fig:model_comparison} shows test accuracy over
communication rounds for SuperSFL, SplitFed, DFL, and HeteroFL
across both datasets, both architectures, and two client scales. SuperSFL converges faster and reaches higher final accuracy than
all three baselines in every setting we tested.
The gap widens with scale: at 100 clients on CIFAR-10, SuperSFL
reaches around 96--98\% while SplitFed plateaus near 70--72\%, and the
confidence bands remain tighter throughout training, suggesting
that TPGF suppresses the instability that grows with the number
of heterogeneous split boundaries. DFL narrows the gap somewhat by adapting split points to resource
profiles, and HeteroFL does so through width scaling, but neither
directly reconciles gradient updates produced at different depths
--- which is what TPGF does.
The advantage holds across both ViT-B/16 and ResNet-18, which
suggests it comes from the optimization mechanism rather than
any architecture-specific property.

Figure~\ref{fig:client-wise-acc} reports the final accuracy achieved by 
individual clients grouped by subnetwork size (Small, Medium, Large). 
As expected, larger subnetworks achieve higher accuracy on average 
(83.4\% $\rightarrow$ 92.8\% $\rightarrow$ 97.1\% for ViT-16 and 
80.5\% $\rightarrow$ 89.7\% $\rightarrow$ 91.9\% for ResNet-18). Importantly, the smallest clients remain competitive despite their limited 
capacity. Although the Small group exhibits larger variance, this variation 
is primarily due to the non-i.i.d.\ data distribution across clients rather 
than training instability. This observation is consistent with the stable 
convergence behavior shown in Figure~\ref{fig:model_comparison}.

\subsection{Communication Efficiency and Time-to-Target-Accuracy}

\begin{table}[t]
\centering
\caption{Communication efficiency comparison: number of rounds, transmitted data volume (MB), and wall-clock time (s) required to reach the target test accuracy under heterogeneous split depths (non-i.i.d. Dirichlet $  \alpha=0.5  $). Bold values indicate the best result in each row. “Cl.” = number of clients; “Tgt.” = target accuracy.}
\label{tab:main_results}
\setlength{\tabcolsep}{4pt}
\renewcommand{\arraystretch}{1.0}
\small
\begin{tabular}{llccrrrr}
\toprule
Dataset & Cl. & Tgt. & Metric 
& SplitFed & DFL & HeteroFL & \textbf{SuperSFL} \\
\midrule
\multirow{6}{*}{CIFAR-10} 
& \multirow{3}{*}{50} & \multirow{3}{*}{80\%}
  & Rounds         & 54      & 39      & 29      & \textbf{13} \\
& & & Comm.\ (MB)  & 61,847  & 43,912  & 71,340  & \textbf{14,280} \\
& & & Time (s)     & 14,210  & 9,840   & 15,120  & \textbf{3,050} \\
\cmidrule(lr){2-8}
& \multirow{3}{*}{100} & \multirow{3}{*}{80\%}
  & Rounds         & 83      & 47      & 36      & \textbf{19} \\
& & & Comm.\ (MB)  & 189,450 & 104,870 & 172,680 & \textbf{42,910} \\
& & & Time (s)     & 21,340  & 11,920  & 18,450  & \textbf{4,680} \\
\midrule
\multirow{6}{*}{CIFAR-100} 
& \multirow{3}{*}{50} & \multirow{3}{*}{75\%}
  & Rounds         & 79      & 52      & 41      & \textbf{23} \\
& & & Comm.\ (MB)  & 91,260  & 59,480  & 99,870  & \textbf{25,940} \\
& & & Time (s)     & 20,180  & 13,110  & 20,940  & \textbf{5,620} \\
\cmidrule(lr){2-8}
& \multirow{3}{*}{100} & \multirow{3}{*}{75\%}
  & Rounds         & 107     & 66      & 54      & \textbf{29} \\
& & & Comm.\ (MB)  & 245,310 & 148,920 & 259,460 & \textbf{65,780} \\
& & & Time (s)     & 27,190  & 16,720  & 27,310  & \textbf{7,180} \\
\bottomrule
\end{tabular}
\end{table}

\begin{figure}[t]
\centering
\includegraphics[width=\textwidth]{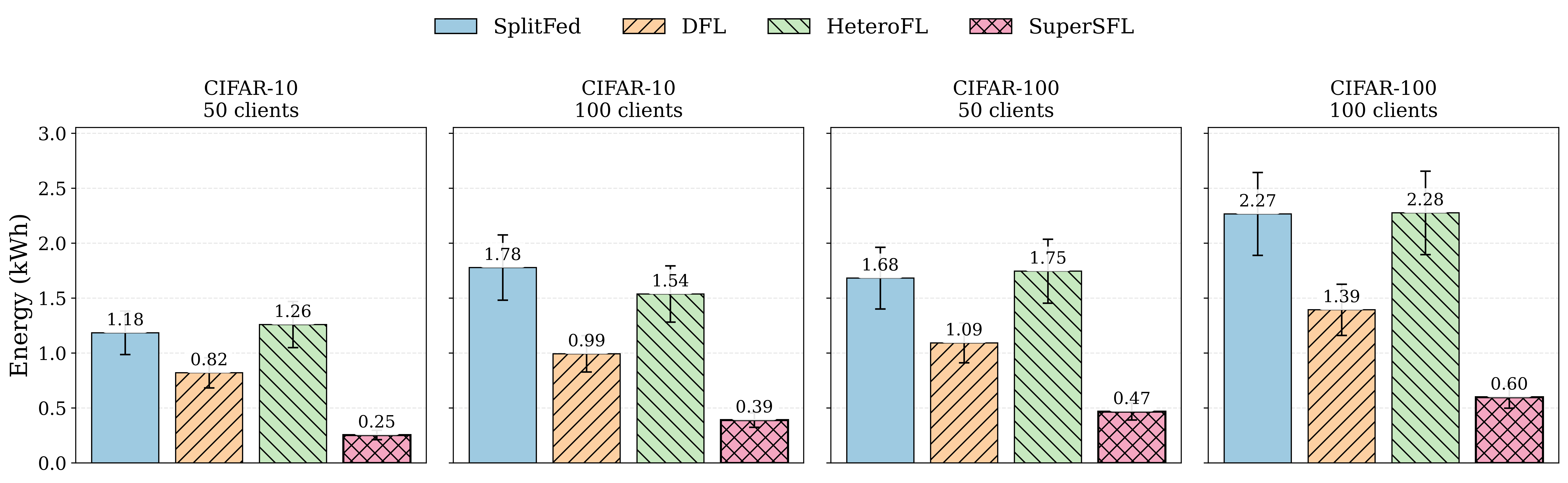}
\caption{Energy consumption (kWh) to reach target accuracy across four settings
(CIFAR-10/100 with 50/100 clients). Error bars denote the GPUs power range.
}
\label{fig:energy}
\vspace{-2em}
\end{figure}

Table~\ref{tab:main_results} compares the cost of reaching a target accuracy 
in terms of communication rounds, transmitted data volume, and wall-clock 
time. SuperSFL consistently achieves the lowest cost across all metrics and 
configurations. SuperSFL requires substantially fewer communication rounds to reach the 
target accuracy. For example, on CIFAR-10 with 100 clients, SuperSFL reaches 
80\% accuracy in 19 rounds, compared with 36 rounds for HeteroFL, 
47 rounds for DFL, and 83 rounds for SplitFed, corresponding to up to a 
$4.4\times$ reduction in training rounds.

The communication savings are even larger when measured in transmitted 
bytes. On CIFAR-10 with 100 clients, SuperSFL transfers 42,910\,MB to reach 
the target accuracy, compared with 172,680\,MB for HeteroFL and 
189,450\,MB for SplitFed, corresponding to around $4.0\times$ reduction 
in communication volume. This improvement arises from both faster 
convergence and lower per-round communication, as shallower client 
prefixes produce smaller activation tensors. Wall-clock time follows the same trend. For instance, on CIFAR-100 with 
100 clients, SuperSFL reaches the target accuracy in 7,180\,s, while 
DFL and SplitFed require 16,720\,s and 27,190\,s respectively, resulting 
in up to a $3.8\times$ speedup. Overall, SuperSFL reduces communication rounds, transmitted data, and 
training time simultaneously, demonstrating that depth-aware split 
coordination can significantly improve system efficiency under 
heterogeneous client configurations.

\subsection{Fault Tolerance under Server Unavailability}

\begin{wraptable}{r}{0.62\columnwidth}
\centering
\caption{SuperSFL test accuracy vs. server-gradient availability. ``Grad (\%)'' = fraction of rounds with server gradients accessible; Mean $\pm$ std over three runs.}
\label{tab:fault_tolerance}
\footnotesize
\setlength{\tabcolsep}{3pt}
\begin{tabular}{>{\centering\arraybackslash}p{0.90cm} p{3.30cm} >{\centering\arraybackslash}p{1.85cm}}
\toprule
\textbf{Grad (\%)} & \textbf{Gradient Source} & \textbf{Acc (\%)} \\
\midrule
100  & Fully server-assisted     & \makebox[1.85cm][c]{95.58 $\pm$ 1.08} \\
70   & Mostly server-assisted    & \makebox[1.85cm][c]{93.81 $\pm$ 2.59} \\
50   & Partially server-assisted & \makebox[1.85cm][c]{93.12 $\pm$ 2.11} \\
20   & Mostly client-driven      & \makebox[1.85cm][c]{91.03 $\pm$ 1.17} \\
10   & Client-driven             & \makebox[1.85cm][c]{89.77 $\pm$ 2.22} \\
0    & Serverless                & \makebox[1.85cm][c]{86.36 $\pm$ 3.25} \\
\bottomrule
\end{tabular}
\end{wraptable}



We evaluate the robustness of SuperSFL under intermittent server availability by varying the fraction of training rounds in which server gradients are accessible. Table~\ref{tab:fault_tolerance} reports the final test accuracy on CIFAR-10 for different levels of server gradient availability. SuperSFL degrades gracefully as server participation decreases. Reducing availability from 100\% to 50\% lowers accuracy by only 2.46 points (95.58\% $\to$ 93.12\%). Even in the fully serverless setting (0\% availability), the model still reaches 86.36\%, showing that meaningful learning can continue without any server support. This robustness comes directly from the lightweight client-side classifier, which lets clients keep training locally during outages; when connectivity returns, those updates integrate naturally into the global supernetwork. We also see a modest increase in variance at lower availability levels (e.g., $\pm$3.25 at 0\% versus $\pm$1.08 at 100\%), which simply reflects the greater stochasticity of purely client-driven updates rather than any instability in the training process.

\subsection{Energy Efficiency}
We measure GPU energy on NVIDIA A100 GPUs using NVIDIA DCGM board-power samples at 100\,ms intervals, following~\cite{patterson2021carbon}. All clients are simulated on the same hardware, and reported energy is accumulated until each method first reaches the target test accuracy. Since A100s differ from the edge devices targeted by SuperSFL, these measurements reflect algorithmic efficiency rather than absolute device-level energy. As shown in figure~\ref{fig:energy}, SuperSFL requires the least energy in all four configurations,
primarily because it reaches the target accuracy in fewer rounds
and transmits smaller activation tensors per round.
On CIFAR-10 with 100 clients it consumes $0.39 \pm 0.03$\,kWh,
compared with $1.78 \pm 0.11$\,kWh for SplitFed and
$1.54 \pm 0.09$\,kWh for HeteroFL.
On CIFAR-100 with 100 clients the gap remains: $0.60 \pm 0.04$\,kWh
for SuperSFL versus $2.27 \pm 0.14$\,kWh for SplitFed.
Across all configurations the reduction ranges from $3\times$ to
$4.6\times$, with larger gains at higher client counts where the
communication savings of shallower prefixes compound over more
rounds.

\subsection{Ablation Study}

\begin{figure}[t]
\centering
\includegraphics[width=\textwidth,height=0.22\textheight,keepaspectratio]{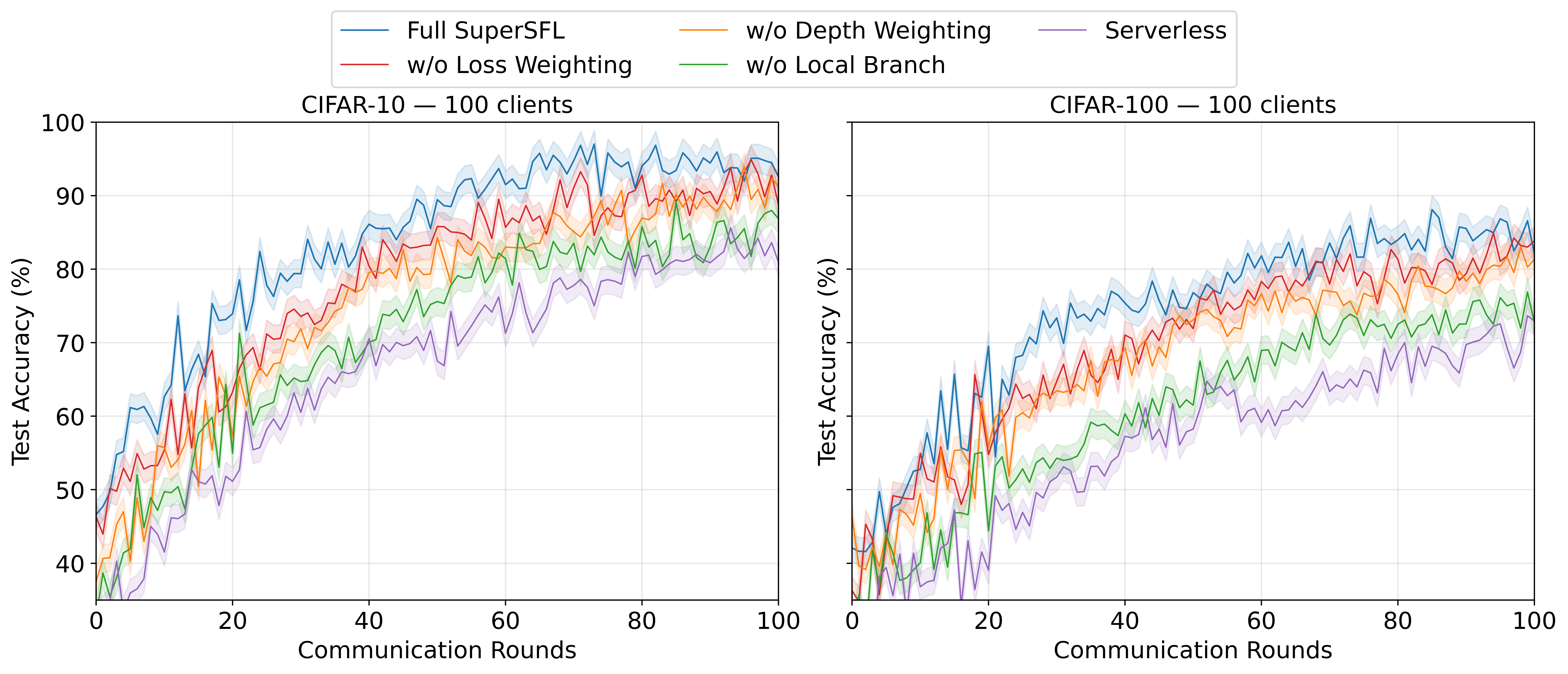}

\caption{Ablation study on CIFAR-10 and CIFAR-100 with 100 clients.
\textit{w/o Loss Weighting} sets $w_\mathrm{client}{=}0.5$
in Eq.~\eqref{eq:direction}; \textit{w/o Depth Weighting} sets
$\rho_i{=}1$ in Eq.~\eqref{eq:coverage}; \textit{w/o Local Branch}
removes the client-side gradient from TPGF; \textit{Serverless}
uses no server gradients.}
\label{fig:ablation}
\vspace{-2em}
\end{figure}

To assess component contributions, we compare full SuperSFL with four controlled variants on CIFAR-10 and CIFAR-100 using 100 clients (Figure~\ref{fig:ablation}). Across both datasets, full SuperSFL achieves the best accuracy and convergence. Removing loss weighting has the smallest effect, disabling the depth-ratio factor causes a larger drop, removing the local branch degrades performance most among partial ablations, and the fully serverless variant performs worst.

The local branch is critical: without it TPGF cannot fuse gradients
and the system falls back to pure server-side supervision, causing
the largest accuracy penalty among the partial ablations, especially
on CIFAR-100.
The two weighting terms are not equally important.
Disabling the coverage scalar $\rho_i$ (Eq.~\eqref{eq:coverage})
produces the larger drop, because it directly distorts how much
influence each client exerts on shared parameters regardless of
how much of the supernetwork it trained.
Removing the inverse-loss direction weight (Eq.~\eqref{eq:direction})
causes a smaller but consistent penalty, reflecting its secondary
role in adjusting trust based on supervision quality.Taken together, these results confirm that the gains of SuperSFL
come from two complementary pieces: the client-side gradient that
makes fusion possible, and the depth- and loss-aware weights that
keep fused updates stable across heterogeneous split boundaries.

\section{Conclusion}
\vspace{-0.7em}

SuperSFL enables heterogeneous clients to train a shared model with different split depths using a weight-sharing supernetwork and the TPGF mechanism. Experiments show faster convergence and reduced communication, time, and energy costs compared with existing methods. However, our evaluation is limited to image classification tasks and simulated environments. Future work will focus on large-scale real-world deployments and extending the framework to other model architectures and learning tasks.

\vspace{-1em}
\subsubsection*{Acknowledgments.}
We would like to thank NSF for  supporting this project (grant 2211982). This work utilized the Delta system at the National Center for Supercomputing Applications (NCSA) through allocation CIS240626. We also acknowledge support from the National Science Foundation under grants 2138259, 2138286, 2138307, 2137603, and 2138296. Additionally, we acknowledge support from Germany's Federal Ministry of Breakthrough Innovation (SPRIN-D) through the Composite Learning Challenge under the SymphonyLearn project.
\subsubsection*{Disclosure of Interests.} 
The authors have no competing interests to declare that are relevant to the content of this article.

\bibliographystyle{splncs04}
\bibliography{reference}

\end{document}